\documentclass%
[prl,amsmath,showpacs,twocolumn,preprintnumbers,superscriptaddress]%
{revtex4}

\usepackage{mathptmx}

\usepackage{microtype}

\usepackage{graphicx}
\usepackage{dcolumn}
\usepackage{amsmath}
\usepackage{mathtools}
\usepackage{bm}

\usepackage{mathptmx}
\usepackage{microtype}

\usepackage{graphicx}
\usepackage{dcolumn}
\usepackage{amsmath}
\usepackage{mathtools}
\usepackage{bm}
\usepackage{amssymb}

\begin{document}

\title{Radiation from charged particles due to explicit symmetry breaking in a gravitational field}

\author{Fabrizio Tamburini} 
\address{ZKM -- Zentrum f\"ur Kunst und Medientechnologie, Lorentzstr. 19, D-76135, Karlsruhe, Germany.}
\address{MSC -- bw,  Stuttgart, Nobelstr. 19, 70569 Stuttgart, Germany.}

\author{Mariafelicia De Laurentis} 
\address{Institute for Theoretical Physics, Goethe University, Max-von-Laue-Str. 1, 60438 Frankfurt, Germany.}
\address{Tomsk State Pedagogical University, ul. Kievskaya, 60, 634061 Tomsk, Russia.}
\address{Lab.Theor.Cosmology,Tomsk State University of Control Systems and Radioelectronics (TUSUR), 634050 Tomsk, Russia}
\address{INFN Sezione  di Napoli, Compl. Univ. di Monte S. Angelo, Edificio G, Via Cinthia, I-80126, Napoli, Italy.}

\author{Ignazio Licata}
\address{Institute for Scientific Methodology (ISEM) Palermo Italy.}
\address{School of Advanced International Studies on Theoretical and Nonlinear Methodologies of Physics, Bari, I-70124, Italy.}
\address{International Institute for Applicable Mathematics and Information Sciences (IIAMIS), B.M. Birla Science Centre, Adarsh Nagar, Hyderabad -- 500 463, India}
 

\begin{abstract}
The paradox of a free falling radiating charged particle in a gravitational field, is a well-known fascinating conceptual challenge that involves classical electrodynamics and general relativity.
We discuss this paradox considering the emission of radiation as a consequence of an explicit space/time symmetry breaking involving the electric field within the trajectory of the particle seen from an external observer. 
This occurs in certain particular cases when the relative motion of the charged particle 
does not follow a geodesic of the motion dictated by the explicit Lagrangian formulation of the problem and thus from the metric of spacetime. The problem is equivalent to the breaking of symmetry within the spatial configuration of a radiating system like an antenna: when the current is not conserved at a certain instant of time within a closed region then emission of radiation occurs \cite{sinha15}. Radiation from a system of charges is possible only when there is explicit breaking of symmetry in the electric field in space and time.
\end {abstract}


\maketitle

\section{Introduction}
From Maxwell's equations, and thus from classical electrodynamics, an accelerated electric point-like charge emits electromagnetic radiation, carrying away energy and momentum \cite{BoBook,jackson,Schwinger}. In this classical representation, the motion of charges generates currents and the acceleration of charges generates electromagnetic waves; this is the mechanism for which an antenna radiates.

Li\'enard--Wiechert potentials provide a comprehensive description of the radiation part of the electric field emitted by a massive accelerated point-like electric charge, expressed in terms of vector and scalar potentials in the Lorenz gauge, which falls off as $1/r$, in addition to its rest-frame Coulomb field that, instead, decays quadratically \cite{landau2,cross}.

Consider now an electric charge falling in a Newtonian gravitational field. From a classical point of view, the Newtonian gravitational field acts as a force onto the particle with the result of accelerating it. In this case, radiation emission from the charged particle is expected. 
Naively, in a \textit{gedankenexperiment} where a neutral and a charged particle are free falling in a Newtonian gravitational field, 
the charged particle should start to radiate, thereby losing energy and slowing down its accelerated motion with respect to the neutral counterpart.

This scenario is conceptually different if described in the framework of General Relativity (GR): gravitation does not act as a classical Newtonian force on the particle but is a manifestation of the spacetime curvature in which the particle is moving. The deformation of the geometry of spacetime is dictated by the local energy and mass distributions that rule the dynamics of any body.
Moreover, GR is built on the constance of the speed of light in vacuum and based on the principle of equivalence between gravitation and inertia and Mach's principle \cite{mtw73}. Being GR a local theory, all our physical experience made through local measurements of events in the spacetime can be reduced to coincidences of point-events \cite{einstein55} with no particular reason for preferring certain systems of coordinates and/or reference frames with respect to other ones. 
From this, the requirement of a general covariance for the laws of physics arises \cite{einstein16} and implies the impossibility of telling through any local measurement the effects of a real gravitational field from a non-inertial system in a flat spacetime. 
This distinction can be made only by causally connecting different events in spacetime. 

By assuming the principle of local equivalence between non-inertial frames and gravitational fields, if the charged particle is expected to radiate in all reference frames, any free-falling observer could be able to violate it locally by using the two-particle experiment: a charged particle in a free-falling laboratory would start to deviate from the neutral particle also in the absence of macroscopic electric fields. Conversely, any charged particle at rest should radiate if its dynamics is described in a non-inertial frame.
From here the paradox of the radiating free falling particle arises.

The assumed universality of the Equivalence Principle (EP) suggests that a particle in a gravitational field has identical physics to one in an accelerated frame.

The best hypothesis that is taken into account to justify the emitted radiation by a static charge in a gravitational field is that of impossibility of finding a real gravitational field, which has an infinite measure exactly corresponding to a real uniformly accelerated motion \cite{bondi55,Unnikrishnan}, such as the Rindler coordinate system \cite{rindler}. In Rindler spacetime a uniformly accelerating particle undergoes hyperbolic motion and there always exist a Rindler frame in which the uniformly accelerated particle is at rest.

The electromagnetic (EM) field of the uniformly accelerated charge behaves as a purely electrostatic field everywhere in the Rindler frame, implying that even observers for which the charge is indeed accelerating when observed from the inertial point of view, would not detect the emitted radiation, proving that the radiation of a uniformly accelerated charge is beyond the horizon \cite{almeida05}.

Moreover, Rohrlich \cite{rohrlich} and Fulton and Rohrlich \cite{fultonrohrlich} proposed appropriate care in distinguishing reference frames for electrodynamics as Maxwell equations should hold only in an inertial frame.
Feynman from energy considerations, instead, claimed that a free falling particle should not radiate at all 
as is expected to occur to a static charge in a gravitational field \cite{feynman}.

When Rindler coordinates are generalized to curved spacetime, they become Fermi normal coordinates, where an appropriate orthonormal tetrad is transported along a given trajectory using the Fermi-Walker transport rule. Their extension for a full description in all the spacetime of the radiating particle paradox seems not fully satisfactory as there are physical phenomena where a uniform gravitational field is not locally equivalent to an accelerating frame because of the broken equivalence between a static gravitational field and an accelerated frame \cite{Unnikrishnan}. 

More complicated argumentations arise also when classical and quantum field effects are invoked. Back-scatter effects of the EM field and other quantum effects have been invoked in the past to discuss more in deep this paradox. As an example, DeWitt \& DeWitt hypothesized that if the charge were moving inertially at non-relativistic velocities in a static gravitational field, the EM field was thought to fall freely with the charge, the net retarding force due to local tidal distortions is zero if integrated over a solid angle. The geodesic deviation of the particle is originated by a field outside the classical radius. Non-local terms arise from a back-scatter process that involves the Coulomb field of the particle and the spacetime deformations and the Coulomb field is shacked back that are propagated back to the particle, where `` the particle tries its best to satisfy the naive EP''  \cite{dewitt64}.
Uniform acceleration of a charged particle facing also possible quantum effects was discussed by Candelas and Sciama \cite{cansci84}, where classical results might appear incorrect either because of the difference of the quantum mechanical equations or because the classical results can give different results for different vacuum states, with an exception for Rindler spacetime \cite{almeida05}.
In any case, distortions of the Coulomb field of the charge due to the curvature tensor of the background gravitational field may give rise to self-forces. 
The problem is that the EP is fully valid for a point-like particle, whilst the field is extended in spacetime and EP cannot always hold for the whole spacetime.

In this paper we will analyze this paradox from a different point of view, by using the fact that a particle should radiate only when an explicit symmetry breaking into the particle dynamics occurs.

\section{The explicit symmetry breaking approach}
Now we discuss a method, based on explicitly broken symmetries, to provide a unified perspective to this problem.
Noether's theorem states that any symmetry corresponds to a conserved physical quantity \cite{noether}.
As an example, the symmetry of electric field lines generated by static charges in a local reference frame and/or in a flat spacetime are associated with the conservation of charges within a localized region of space and time.
Point-like particles at rest provide for rotational symmetry of the Lagrangian and, because of this, the Lagrangian has angular invariance. Accelerated charges in the Rindler spacetime, instead, exhibit translational symmetry with respect to the direction of the hyperbolic motion and in this case the Lagrangian shows translational symmetry across the direction of the uniformly accelerated motion.

Symmetries of a system can be also put in evidence through analogies with other physical systems. 
Following the approach by Landau and Lifshitz \cite{landau2}, the equations of electrodynamics in the presence of a gravitational field can be mathematically described by equations of electrodynamics in dielectric media and mimic the effect of a gravitational field following the prescriptions of analogue gravity. This approach investigates analogues of general relativistic gravitational fields within other physical systems \cite{barc05}. 
The problem of the radiation from point-like electric charged particles in a gravitational field can be discussed in terms of analog gravity and through the equivalence between particles moving in a gravitational fields and charges in motion across different trajectories and moving through a dielectric.
In this case there is radiation when symmetry is explicitly broken, like in asymmetric resonators such as the antenna invented by Marconi in the pioneering days of radio communication or dielectric resonator antennas \cite{marconi}, where the symmetry breaking is correlated to the temporal variance of its Lagrangian and loss of Noether current from the system.
This means that the charge radiates when the trajectory of the point-like charged particle is not a geodesic and the derivatives of the Lagrangian with respect to the four coordinates are not null.
Of course because of EP this is valid only in a point, because any point that does not belong to the geodesic under consideration will experience tidal forces, seen as a breaking of symmetry due to the terms from the equations of the geodesic deviation.

In practice, the symmetry breaking corresponds to the Larmor's "whiplash".
The problem of Feynman's approach is that the emission of radiation is due to a variation in the acceleration. 
A constant acceleration does not produce radiation. 
In support of its argument, there would be the reaction force of Abraham-Lorentz in which appears the derivative of acceleration. 
In a hypothetical uniform gravitational field there is not a problem because the charge does not radiate, being the acceleration that of a constant fall. 

Moreover, a charge at rest in a gravitational field is equivalent to an "uniformly accelerated" charge and then it does not radiate. In this way the EP seems heuristically saved. 
But, the real problem are the constraints imposed by the mathematics of GR. In fact all the differential relations valid in special relativity are transcribed in GR with the principle of minimal coupling by replacing the partial derivatives with covariant ones. 
Here, the density of point-like charges $e_a$ and the correspondent currents are described as follows,
\begin{eqnarray}
&& \rho = \sum_a \frac{e_a}{\sqrt{\gamma}} \delta(\mathbf{r}-\mathbf{r}_a)\,,
\\
&& j^i = \sum_a \frac{e_a c}{\sqrt{-g}} \delta(\mathbf{r}-\mathbf{r}_a) \frac{d x^i}{d x^0}\,,
\end{eqnarray}
where $\gamma$ is the metric tensor determinant, $\mathbf{r}_a$ the position of the $a-$th electric charge. The current conservation law becomes
\begin{equation}
j_{;i}^i = \frac 1{\sqrt{-g}} \frac{\partial}{\partial x^i}\left(\sqrt{-g} j^i\right) = 0\,,
\end{equation}
this equivalence with a dielectric becomes more evident when one writes the equations in a three-dimensional formalism in a gravitational field with the determinant $\gamma$ that varies in time.

In the relativistic Larmor formula \cite{jackson} when the partial derivatives are replaced with the covariant ones, we get that the charge radiates if and only if the four acceleration is nonzero. But this means that the charge radiates if and only if it is not on a geodesic of that given spacetime.
 
One can conclude that the point-like charge that falls in the real field is located on a geodesic and therefore should not radiate. Feynman's solution thus preserves the EP but at all facts violates the deep formal structure of the GR that the EP supports, and is valid only in case of a point-like source where tidal forces do not apply.

From classical EM field theory and antenna theory, a particle radiates in presence of explicit symmetry breaking  \cite{sinha15}. 
Explicit symmetry breaking is associated with a condition where the dynamic equations and the Lagrangian of the system are not invariant due to some terms that break its symmetry \cite{castellani}.
This means that the particle is forced away from the geodesic motion.
Radiation from a system of accelerating charges occurs only in the presence of an explicit symmetry breaking, namely, a symmetry breaking ``in the electric field in space within the spatial configuration of the radiating system''. 
What happens is that, even if from Maxwell's equations currents and charges are globally conserved, when an explicit symmetry breaking is involved, currents around a closed area in the neighborhood of the radiating structure appear not to be locally conserved at a certain instant of time, with the result of emitting radiation.
If no external EM field is present, the Lagrangian of the system is the classical Einstein's Lagrangian,
\begin{equation}
S_g = \int G \sqrt{-g} d \Omega\,,
\end{equation}
instead, if EM field is considered, the Lagrangian is
\begin{equation}
S =  \int G \sqrt{-g} d \Omega - \frac 1{16 \pi} \int g^{ir} g^{js} F_{ij} F_{rs} d \Omega\,.
\end{equation}
The condition $\delta S = 0$ gives the geodesic equation. 
The symmetries of the Lagrangian depend on the structure of spacetime and are expressed in terms of translational, rotational and other Noether invariants. The main symmetry derived from the Lagrangian in the case of  the motion of a point-like falling charge is expressed by its geodesic motion.
The symmetry related to the EM field is preserved when the derivatives of the Lagrangian with respect to the coordinate system is null. 
Explicit symmetry breaking occurs when the particle is forced in a trajectory that does not follow the geodesic motion, forced to move into paths that do not directly satisfy the least action principle. When charged particles are forced onto these paths they are expected to radiate as recently shown \cite{sinha15} for classical and dielectric antennas.

Similarly this can happen in GR. The motion of a charged particle in a given spacetime can be described in the same terms as a charge inside an antenna, by using the analogies between gravitation and dielectrics for electromagnetic phenomena.
In a static gravitational field such as in Schwarzschild spacetime a free falling particle does not radiate, being mathematically equivalent to a dielectric with electric and magnetic permittivity $\epsilon = \mu = 1 / \sqrt{h}$, where $h$ is the three-dimensional spatial metric tensor determinant.

In the context of static charges, we can easily associate the symmetry of electric field lines with the conservation of charges within a localized region of space and time.
The affine connection terms give the relative acceleration equivalent term from one geodesic to another.
What happens is that when the charge is forced to a periodic acceleration, the symmetry of the electric field is explicitly broken within a localized region of space and time resulting in rotation of the electric field which generates magnetic field resulting in emission of electromagnetic radiation from the charge center.

\subsection{Antennas, geodesics and Lagrangians}

The problem of a radiating particle in GR finds an interesting parallelism with an analogue gravity approach to GR and antenna theory.
An antenna radiates when the field lines present a broken symmetry or, equivalenlty, when the charges in the antenna have an accelerated motion.
The charges in antennas emitting monochromatic radiation are animated by a particular motion, harmonic motion.
Noteworthy, an antenna has always a finite extension in space and charges oscillate in time \cite{reviewpaper}.

Let us describe the motion of charges as in a radiating antenna in terms of space and time symmetries of the electric field within the spatial configuration of the radiating system as experimentally proved1\ \cite{sinha15} and apply to our problem in GR.

Radiation from a system of accelerating charges is possible only when there is explicit breaking of symmetry in the electric field in space within the spatial configuration of the radiating system.
An antenna works by explicitly breaking the symmetry of the transmission line for the electric field.
Explicit symmetry breaking is associated with a condition where the dynamic equations and the Lagrangian of the system are not invariant due to some terms that break its symmetry and some physical quantity must lose its conserved value within a localized region of space and time.

Charges radiate when external fields and accelerations deviate the charge from the geodesic motion described by the Lagrangian and not always is possible to separate space and time in the whole manifold. 
This happens also when EM backreaction is considered.
In the classical representation, gravitational fields without a timelike Killing vector do not have an invariant definition of ``at rest'' state and the self-force cannot be invariantly decomposed into radiation reaction and gravitationally-induced self-force.
Maxwell equations are conformal invariant under a transformation of this type $\tilde g_{ik} = \Omega^2 g_{ik}$
which implies that $\tilde F^{ij} = \Omega^{-4} F^{ij}$ and $\sqrt{-\tilde g} = \Omega^4 \sqrt{- g}$ and the invariance of Maxwell equations. 
The equation of motion for a charged test particle in a generic Riemannian spacetime in an external EM field $F^{ext}$ is
\cite{defelice}
\begin{eqnarray}
m_0 c \frac{D u^i}{D \tau} &=& \frac ec \left( F^{ext} \right)^i_k u^k +     
\\
&+& \frac 23 \frac{e^2}{c^3} \left[ \frac{D^2 u^i}{D \tau^2} - \frac{u^i}{c^2} \left( \frac{D u^k}{D \tau} \frac{D u_k}{D \tau} \right) \right] +  \nonumber
\\
&-& \frac{e^2}{3c} \left( R^i_j u^i + \frac {1}{c^2} u^i R_{jk} u^i u^k \right) +    \nonumber
\\
&+& \frac{e^2}{c} u^k \int_{- \infty}^\tau f^i_{kj} u^j (\tau ') d \tau ' \,,   \nonumber
\label{irradia}
\end{eqnarray}
Any charged particle in a general spacetime, also in the absence of an external EM field ($F^{ext}=0$) will deviate from the geodesics of a neutral particle because of the radiation damping, written in the second line of the equation, the Hobbs term and the non-local contribution of the history of the particle on the curved spacetime.
This implies emission of radiation.
As is well known, not to radiate, a particle has to have all the four terms equal to zero: null 4-velocity, null radiation damping or up to a constant with a Rindler spacetime-like behavior and null curvature, namely, $R_{ik}=0$.
If one considers a field around the charge, the field backreaction should always induce radiation as it behaves as an extended object and experiences geodesic deviation.

In any case we find conceptual limits to a clear and unique definition to this problem.
The problem of a radiating particle in GR using classical concepts seems to be an ill-posed problem, but for certain particular cases described by Eq. \ref{irradia}. It is not possible to state in general whether a point-like charged particle is radiating for any observer in the whole spacetime.

Conversely, if we consider the actual properties of the mathematical structure of GR, theory connects an event in spacetime to another, in a point-to-point correspondence of a given manifold. GR is made with coincidences of events in/of spacetime.
This implies that there always exist a reference frame, the one coincident with the particle and its motion, where the particle is not radiating at all. 
Thus, in the strict language of GR, without considering the field around the charge as additional hypothesis, one finds a contradictory statement. A particle in accelerated motion should always radiate, but there is always a reference frame where the particle does not radiate.

This appears to be contradictory, because, in any case, from a classical point of view, to emit/detect radiation we need space and time intervals: space for the oscillation of the electric field, space intervals to host the motion of charges that emits a precise wavelength like in an antenna, or time intervals to emit/detect locally the oscillations of the electric field.
Classical radiation implies space and time foliation and space and time intervals, while GR and EP appear to be in contrast with these needs, implying point-to-point coincidences of events in spacetime.

An agreement with the claims by Rohrlich \cite{rohrlich} and Fulton and Rohrlich \cite{fultonrohrlich} is found in spacetimes where a foliation is possible, a situation that cannot be always obtained in GR as general solution for any spacetime in the whole spacetime. In fact, foliation, the separation of spacetime in space and time, is always possible locally, but a global space-like foliation where antennas and currents can radiate may not even exist in particular spacetimes, because, in general, the metric tensor of a spacetime has no symmetries. Only in particular cases of metric tensors with peculiar symmetries is possible to slice spacetime in a preferred way following these symmetries \cite{Stachel}.

Thus, in a ``G\"odel way'', the sentence ``It is always possible to determine if an accelerated point charge radiates or not in any spacetime for any observer'' seems to be contradictory if one considers the classical aspects of GR and electromagnetism.

\subsection{Emergent gravity approach}
Finally, we briefly discuss this problem if gravitation were an emerging force. We notice that also with the emergent gravity approach \cite{verlinde11,verlinde16}  one finds the same results for the problem of the falling charged particle.
Unruh showed that an observer in an accelerated frame experiences a temperature
\begin{equation}
\label{unruh}
 k_BT= {1\over 2\pi} {\hbar a\over  c},
\end{equation}
where  $a$ denotes the acceleration.

If gravity is emergent, Uhnru equation (\ref{unruh}) does not give the temperature caused by an acceleration and connected to a possible radiation emitted by an accelerated charge, instead it should be interpreted as a formula for the  temperature $T$  required to cause an acceleration equal to $a$.
In order to have a non zero force, we need to have a non vanishing temperature. 

Consider the force that acts on a particle of mass $m$. In a general relativistic setting force can be transformed  by a general coordinate transformation. By using the time-like Killing vector one can give an invariant meaning to the concept of force, needing a foliation of spacetime as previously discussed.
To briefly introduce this alternative approach, we consider an holographic screen $\cal S$ corresponding to an equipotential surface of a fixed gravitational potential.
The four velocity $u^a$ of the particle and its acceleration  $a^b\!\equiv\!u^a\nabla_a u^b$  can be expressed in terms of the Killing vector $\xi^b$ as
$$
u^b =e^{-\phi}\xi^b, \qquad\qquad a^b=e^{-2\phi}\xi^a\nabla_a\xi^b.
$$
And, from the Killing equation one obtains
$$
\nabla_a\xi_b+\nabla_b\xi_a=0\,,
$$ 
and provides the definition of a potential $\phi$.  One finds that the acceleration can again be simply expressed the gradient
\begin{equation}
\label{relaccel}
a^b  =-\nabla^b\phi.
\end{equation}
Note that the acceleration is perpendicular to $\cal S$. Because of this, we can write it as a scalar through an index contraction with a unit vector pointing outwards $N^b$ orthogonal to $\cal S$ and to the Killing vector $\xi^b$. 
This suggests that there is no radiation emitted when the variation of entropy is zero, namely when adiabatic or better isentropic lines that correspond to the geodesic lines of motion are considered.

\section{Conclusions}
General Relativity and classical electromagnetism are two classical theories, still incomplete to give a complete and unique description to the classical problem of the radiation emitted from a point-like charge in a gravitational field.
In certain particular cases  the problem can be faced and solved by using the symmetries of the metric tensor that allows a space time foliation of the four dimensional spacetime and one can describe the radiation of a charge in motion in terms of symmetry breaking of the Lagrangian of the system, that correspond to non-geodesic motion of the point-like charge in the gravitational field.
In general, instead, the solution to the problem is not unique and clear because the problem is not well posed: either one limits the analysis to a point-like charge and the equivalence principle and finds a family of reference frames where the charge never radiates, according to Feynman assumptions, or, by including the interaction and backreaction of the particle with the EM field or, including also quantum effects, finds geodesic deviations and emission of radiation because EP cannot always hold in a neighborhood of the charge to be equivalent to a Rindler spacetime.
In conclusion, the problem of a moving and falling charge in a gravitational field - as it is - does not have a general solution. GR is mathematically written in terms of correspondence between point like events that build spacetime, whilst the emission of radiation of a charged particle seen from a classical EM point of view needs a local spacetime where there are intervals of space and time needed to generate and measure the oscillations of the EM radiation, namely it requires a foliation of the spacetime which is not strictly local and that exists only when the metric tensor of spacetime presents certain symmetries. Emergent gravity seems to face the same problems that will be presumably solved in a future quantum gravity description.

\section*{Acknowledgements}
M.D.L.\ is supported by Grant "BlackHoleCam" Imaging the Event Horizon of Black Holes awarded by the ERC in 2013 (Grant No. 610058). M.D.L acknowledges the COST Action CA15117 (CANTATA) and INFN Sez. di Napoli (Iniziative Specifiche QGSKY and TEONGRAV).


\begin{thebibliography}{ }
\bibitem{sinha15} 
D. Sinha, G. A. J. Amaratunga, {\it Phys. Rev. Lett.}, \textbf{114},  147701(7) (2015).
\bibitem{BoBook}
B. Thid\'e, \textit{Electromagnetic Field Theory}, 2nd ed. (Dover Publications, Inc., Mineola, NY, USA, 2011 \textit{In press}).

\bibitem{jackson} 
J. D. Jackson, \textit{Classical Electrodynamics}, 3rd ed. (Wiley \& Sons, New York, NY, USA, 1998).

\bibitem{Schwinger} 
J. Schwinger, L.L.Jr. DeRaad, K.A. Milton, T. Wu-yang, \textit{Classical Electrodynamics}, (Perseus Books, Reading, MA, USA 1998).

\bibitem{cross}
D. J. Cross, Am. J. of Physics \textbf{83} (4): 349-352 (2015).

\bibitem{landau2} 
 L. D. Landau   and E.M. Lifshitz, \textit{The Classical Theory of Fields}, (Course of Theoretical Physics, Vol. 2, Pergamon, Oxford, UK, 1975).

\bibitem{mtw73}
C. W. Misner, K.S. Thorne, \& J.A. Wheeler, \textit{Gravitation}, (United States: W H Freeman and Company, San Francisco, USA, 1973).

\bibitem{einstein55}
A. Einstein, ``Time, space and gravitation (1948)''  in \textit{``Out of my later years''} 54-58, New York , Philosophical Library (1950).

\bibitem{einstein16}
A. Einstein, {\it Ann. Phys.} (Leipzig), \textbf{49}, 769--822 (1916).

\bibitem{bondi55}
H. Bondi,  and T. Gold, {\it Proc. Royal Soc. London} \textbf{229}, 416--424 (1955).

\bibitem{Unnikrishnan}
C. S. Unnikrishnan, G. T. Gillies, {\it 	Int. J. Mod. Phys. D} {\bf 23}, 1442008 (2014). 

\bibitem{rindler}
W. Rindler, \textit{Essential Relativity}. New York, Van Nostrand Reinhold Co. NY, USA (1969).

\bibitem{almeida05}
C. de Almeida, and A. Saa,  {\it AmJPhys} \textbf{74}, 154--158 (2006).

\bibitem{rohrlich} 
F. Rohrlich, \textit{Classical Charged Particles}, (World Scientific Publishing, Singapore 2007).

\bibitem{fultonrohrlich}
T. Fulton,  and F. Rohrlich,  {\it Annals of Physics} \textbf{9}, 499--517 (1960).

\bibitem{feynman}
 R. P. Feynman, \textit{Feynman Lectures on Gravitation}, lecture 9, (B. Hatfield, Ed., Addison-Wesley, MA, 1995).



\bibitem{dewitt64} 
C. M. DeWitt  and B.S. DeWitt,  {\it Physics} \textbf{1}, 3 (1964).

\bibitem{cansci84}
P. Candelas, and D. W. Sciama, in \textit{Quantum Theory of Gravity} Essays in honor of the 60th birthday of Bryce S. DeWitt, Christensen, S. M., ed., Adam Hilger Ltd, Bristol, UK (1984).

\bibitem{noether}
E. Noether,  {\it Nachr. Ges. Wiss. Goettingen Math.-Phys. Kl. }\textbf{325}, 235. English transl.: Invariant variation problems, Transp. Theor. Stat. Phys., 1, 186--207 (1918).

\bibitem{barc05}
C. Barcelo, S. Liberati, and M. Visser, {\it Living Rev. Relativity}, {\bf 14}, 3, http://www.livingreviews.org/lrr-2011-3  (2011).

\bibitem{marconi}G. Marconi  {\it Proc. IRE},  {\bf 10}, 215?238, (1922).

\bibitem{castellani}
K. Brading,  and E. Castellani, {\it Symmetries in Physics: Philosophical Reflections} (Cambridge University Press, Cambridge, 2003).

\bibitem{reviewpaper}
B. Thid/'e, F. Tamburini, H. Then, C.G. Someda, R.A. Ravanelli,  The physics of angular momentum radio, arXiv preprint arXiv:1410.4268 (2014).
 
\bibitem{defelice}
F. de Felice, and C. j. S. Clarke, \textit{Relativity on Curved Manifolds}, (Cambridge University Press - NY, USA 1990).

\bibitem{Stachel}
J. Stachel, Develpoment of the Concepts of Space, Time and Space-Time from Newton to Einstein, in \textit{100 Years of Relativity: Space-Time Structure: Einstein and Beyond}, (Ed. Abhay Ashtekar, World Scientific, Singapore 2005).

\bibitem{verlinde11}
E. P. Verlinde, {\it JHEP} {\bf 1104}, 029 (2011).

\bibitem{verlinde16}
E. P. Verlinde,  {\it Emergent Gravity and the Dark Universe} ArXiv preprint arXiv:1611.02269 [hep-th] (2016).

\end{thebibliography}
\end{document}